\documentclass[doublecol]{epl2} 
\usepackage{graphicx}
\usepackage{bm}
\usepackage{color}
\usepackage{lipsum}
\usepackage{soul}
\usepackage{mathtools}
\usepackage{bbm}
\usepackage{url}
\usepackage{MnSymbol}

\title{Thermodynamic constraints on kinetic perturbations of homogeneous driven diffusions}
\shorttitle{Diffusive kinetic perturbations} 

\author{Qi Gao\inst{1} \and Hyun-Myung Chun\inst{2} \and Jordan M. Horowitz\inst{1,3,4}}
\shortauthor{Qi Gao \etal}

\institute{                    
  \inst{1} Department of Physics, University of Michigan, Ann Arbor, Michigan, 48109, USA\\
  \inst{2} School of Physics, Korea Institute for Advanced Study, Seoul, South Korea\\
  \inst{3} Department of Biophysics, University of Michigan, Ann Arbor, Michigan, 48109, USA\\
  \inst{4} Center for the Study of Complex Systems, University of Michigan, Ann Arbor, Michigan 48104, USA
}

\abstract{
We analyze the static response to kinetic perturbations of nonequilibrium steady states that can be modeled as diffusions.
We demonstrate that kinetic response is purely a nonequilibirum effect, measuring the degree to which the Fluctuation-Dissipation Theorem is violated out of equilibrium.
For driven diffusions in a flat landscape, we further demonstrate that such response is constrained by the strength of the nonequilibrium driving via quantitative inequalities.}
\begin{document}
\maketitle
\section{Introduction} The question addressed by linear response theory is how a system reacts to a small perturbation~\cite{Kubo,MARCONI2008111}.
Traditionally the only perturbation considered was the application of a small force.
The reason is that around equilibrium steady states, changes in kinetic parameters---like the mobility of a colloidal particle or the energy-barrier along a reaction pathway---are trivial in that they have no effect: the  Boltzmann distribution only depends on the energies of the system's states.
Out of equilibrium, this is not the case, and kinetic perturbations not only affect the steady-state distribution, but including their effects are necessary to completely capture nonequilibrium response.

We are learning now that explicitly analyzing kinetic perturbations can lead to quantifiable insight into nonequilibrium response.  
Indeed, the extent to which the Fluctuation-Dissipation Theorem (FDT) is broken out of equilibrium equals a kinetic response \cite{PhysRevX.10.011066,PhysRevE.105.L012102}.
In addition, for Markov jump processes~\cite{PhysRevX.10.011066,Martins2023,Aslyamov2024}, chemical reaction networks~\cite{Chun2023}, and one-dimensional diffusions~\cite{PhysRevE.105.L012102}, we can put concrete limits (or bounds) on the kinetic response.  It has not been shown, however, that the response remains bounded for driven diffusions away from equilibrium in dimensions higher than one. Here, we first highlight the fact that kinetic perturbations in arbitrary diffusion processes measure the degree to which the FDT is broken.
Then for homogeneous diffusions in a flat landscape we derive bounds on the kinetic response in terms of the strength of nonequilibrium driving.
These results demonstrate that at least for this class of diffusions, response is indeed bounded, and allow us to speculate that similar limits may hold in general for arbitrary diffusions.

This work complements other approaches to rationalizing nonequilibrium response.
A number of such predictions have been inspired by the FDT~\cite{Baiesi2013,Baldovin2022}, and focus on linking response to  correlation functions with the nonequilibrium potential~\cite{Agarwal1972,Prost2009}, stochastic entropy production~\cite{Seifert2010}, dynamical activity~\cite{Baiesi2009}, or force \cite{Caprini2021}. Alternatively, one can analyze the violation of the equilibrium FDT by introducing an effective temperature \cite{BenIsaac2011,Cugliandolo2011,Dieterich2015} or connecting the violation to entropy production via the Harada-Sasa equality \cite{PhysRevLett.95.130602,Toyabe2010,Lippiello2014,Wang2016}. 
As bounds, our predictions allow for simple interpretations at the cost of replacing equalities with inequalities.
From this point of view, our work fits into a growing literature aiming to understand nonequilibrium behavior by placing thermodynamic limits on observable phenomenology~\cite{Baiesi2011,Barato2015,Gingrich2016, Barato2017, Uhl2019, Horowitz2020,Dechant6430,Liang2022, Oberreiter2022, Owen2023b, Liang2023, Dechant2023, Arunachalam2023}.

\section{Setup} We have in mind $D$-dimensional systems whose configurations evolve with time according to a periodic diffusion process.  
For this class of systems, their configuration ${\bf x}(t) = (x_1(t),\dots x_D(t))$ at time $t$ takes values in the torus $\Omega = [0,L]\times\cdots\times[0,L]$, and the Fokker-Planck equation describing the time-evolution of the probability density $p({\bf x},t)$ can be parameterized as~\cite{Gardiner}
\begin{align}
\nonumber
\partial_t p({\bf x},t) &= -\nabla\cdot [{\hat \mu}({\bf x})\cdot (-\nabla U({\bf x})+{\bf F}({\bf x}))p({\bf x},t)]\\  \label{eq:FPE}
&\qquad\qquad\qquad\qquad\qquad+\nabla\cdot{\hat\mu}({\bf x})\cdot\nabla p({\bf x},t)\\ 
&\equiv{\mathcal L}p({\bf x},t).\label{eq:FPO}
\end{align}
Borrowing language from the modeling of a colloidal particle in a viscous fluid, we have identified in the Fokker-Planck operator  ${\mathcal L}$ a positive-definite, mobility matrix ${\hat\mu}({\bf x})$ as well as having split the force into a conservative part due to a potential $U({\bf x})$ and a nonconservative part ${\bf F}({\bf x})$ ($\nabla\times{\bf F}\neq 0$).
We will assume, that the system relaxes to a unique steady-state distribution $\pi({\bf x})$ given as the solution of ${\mathcal L}\pi({\bf x})=0$.
In general, this steady-state solution is not known.  
However, when the nonconservative force is zero (${\bf F}=0$),  it is straightforward to show that the steady-state distribution is $\pi^{\rm eq}({\bf x}) \propto e^{-U({\bf x})}$ (with $k_BT=1$), which we will identify as an equilibrium distribution.

\section{Linear response} A central paradigm in statistical physics is to analyze a system by how steady-state averages of observables, $\langle Q\rangle = \int Q({\bf x})\pi({\bf x})d{\bf x}$, change in response to external perturbations.

Response has traditionally been modeled by assuming that the potential depends on an external parameter $\lambda$ via $U({\bf x}) \to U({\bf x})-\lambda V({\bf x})$, where we call the function $V({\bf x})$ the conjugate coordinate.
In response, averages $\langle Q\rangle$ change by~\cite{MARCONI2008111}
\begin{equation}
R_U =\partial_\lambda \langle Q\rangle= \int V({\bf z})\frac{\delta\langle Q\rangle}{\delta U({\bf z})}d{\bf z}.
\end{equation}
For equilibrium systems, where ${\bf F}=0$, the response is completely captured by how the equilibrium  distribution $\pi^{\rm eq}({\bf x};\lambda) \propto e^{-(U({\bf x})-\lambda V({\bf x}))}$ is modified.
The immediate consequence of this structure is the FDT, relating the static response of the equilibrium average of the observable $\langle Q\rangle^{\rm{eq}} = \int Q({\bf x})\pi^{\rm{eq}}({\bf x})d{\bf x}$ to the fluctuations~\cite{MARCONI2008111}
\begin{equation}\label{eq:FDT}
R_U^{\rm eq} = \int V({\bf z})\frac{\delta\langle Q\rangle^{\rm eq}}{\delta U({\bf z})}d{\bf z}= \llangle Q,V\rrangle^{\rm eq},
\end{equation}
where the covariance is $\llangle Q,V\rrangle^{\rm eq} = \langle Q V\rangle^{\rm eq} - \langle Q\rangle^{\rm eq}\langle V\rangle^{\rm eq}$.  
Perturbations of the kinetics via the mobility ${\hat \mu}({\bf x})$, by contrast, have no effect as it does not enter the equilibrium  distribution.

Now, our previous analyses \cite{PhysRevX.10.011066,PhysRevE.105.L012102,Chun2023, Martins2023} have revealed that away from thermodynamic equilibrium it is in fact useful to consider how observables change in response to perturbations of the kinetics.
This is implemented by allowing the mobility to depend on the external parameter instead, ${\hat \mu}({\bf x}) \to {\hat \mu}({\bf x}) (1-\lambda V({\bf x}))$,
\begin{equation}
R_\mu = \int V({\bf z})\sum_{i,j=1}^N\frac{\delta \langle Q\rangle}{\delta \ln \mu_{ij}({\bf z})}d{\bf z}.
\end{equation}
Applying this perturbation in an experimental setting, where ${\hat \mu}$ is the only system parameter varied, is likely challenging but may be possible.
For example, the mobility of a colloidal particle depends on the viscosity of the surrounding fluid via the Stokes-Einstein relation.
One could then imagine varying the mobility by mixing fluids of differing viscosities.
Nevertheless, kinetic perturbations serve as an important intermediary in our analysis of energy perturbations  \eqref{eq:FDT}  away from equilibrium. 
Here, they serve as a measure of the violation of the FDT \cite{Graham1977, Chun2021}
\begin{equation}\label{eq:viol_FDT}
R_\mu =R_U- \llangle Q,V\rrangle,
\end{equation}
with $\llangle Q,V\rrangle$ the nonequilibrium covariance.
Now, $R_U$ and $\llangle Q,V\rrangle$ are commonly measured in experiments.
Their difference is a purely nonequilibrium effect captured by the kinetic response.

One might then surmise that a larger kinetic response requires stronger nonequilibrium driving.
Indeed, our previous work has shown that there are such quantitative trade-offs, at least for one-dimensional diffusions on the circle.
 In this case, the only nontrivial form for the nonconservative force is a constant $f$.
Then we have shown that when $V(x) = \delta_{(a,b)}(x)$ is an indicator function on the range $x\in (a,b)$ and choosing $Q(x)\in [0,1]$ for ease of presentation,  the response to a kinetic perturbation can be bounded by the strength of the nonequilibrium driving~\cite{PhysRevE.105.L012102}
\begin{equation}\label{eq:1Dbound}
\left|\int_{a}^b \frac{\delta \langle Q\rangle }{\delta \ln \mu(z)}dz\right| \le \langle Q\rangle (1-\langle Q\rangle)\tanh(|f|L/4).
\end{equation}
The product $\langle Q\rangle(1-\langle Q\rangle)$ represents the maximum variance in the observable possible with fixed mean via the Bahtia-Davis inequality $\llangle Q^2\rrangle \le \langle Q\rangle(1-\langle Q\rangle)$~\cite{Hossjer2022}.
Thus, \eqref{eq:1Dbound} can be viewed as a trade-off between the response, fluctuations, and thermodynamic driving.
Viewed another way,  \eqref{eq:1Dbound} is the maximum of the kinetic response over the system parameters, $\mu(x)$ and $U(x)$, holding fixed the thermodynamic driving $f$ and the observable's average $\langle Q\rangle$.

The derivation of \eqref{eq:1Dbound} relied on having a closed-form analytic solution for the steady-state distribution in one dimension for arbitrary system parameters.
Such a method does not translate to diffusions in higher dimensions where no such solution is known.  
This naturally raises the question of whether thermodynamic force is a constraint on the kinetic response for diffusions in higher dimensions, or if unbounded response is possible?
We address this question in the next section.
\section{Bounds on kinetic response} Without a closed-form solution to the Fokker-Planck equation \eqref{eq:FPE} in dimensions higher than one, we make progress by focusing on a simpler class of models: homogenous driven diffusions in a flat landscape.  
Specifically, for the remainder of this article we specialize to the case where the mobility matrix is constant and diagonal, ${\hat\mu}({\bf x}) = {\hat \mu}$ with diagonal elements $\{\mu_1,\dots,\mu_D\}$; there is no potential $U({\bf x})=0$; and the nonconservative driving is uniform  ${\bf F}({\bf x}) = {\bf f} = (f_1,\dots, f_D)$.
In this case, the dynamics of the probability distribution is determined by the Fokker-Planck operator \eqref{eq:FPO}
\begin{equation}
{\mathcal L} = -\nabla\cdot {\hat \mu}\cdot {\bf f}+\nabla\cdot{\hat\mu}\cdot\nabla,
\end{equation}
with constant coefficients.
The translational-symmetry implies that the steady-state solution is uniform, $\pi({\bf x}) = 1/|\Omega|$, which can also be verified by direct substitution.
Despite the simplicity of the steady-state distribution, spatially-dependent perturbations of the mobility can still lead to complicated changes in the steady-state distribution.
It is these responses that we aim to constrain.

For kinetic perturbations of driven diffusions in a flat landscape, we have derived three thermodynamic limits. In contrast to our previous work \cite{PhysRevE.105.L012102}, 
we cannot optimize over all system parameters, as they are fixed.
Instead, we maximize the response over the observable $Q(x)$ and conjugate coordinate $V(x)$.
To have a well-posed problem, though, we need to constrain $Q(x)$ and $V(x)$ in some way.
In light of our previous results \eqref{eq:1Dbound} and taking inspiration from the FDT, we fix their fluctuations via their variances, $\llangle Q^2\rrangle$ and $\llangle V^2\rrangle$.
 
\begin{figure}[t]
\begin{center}
\includegraphics[width=.45\textwidth]{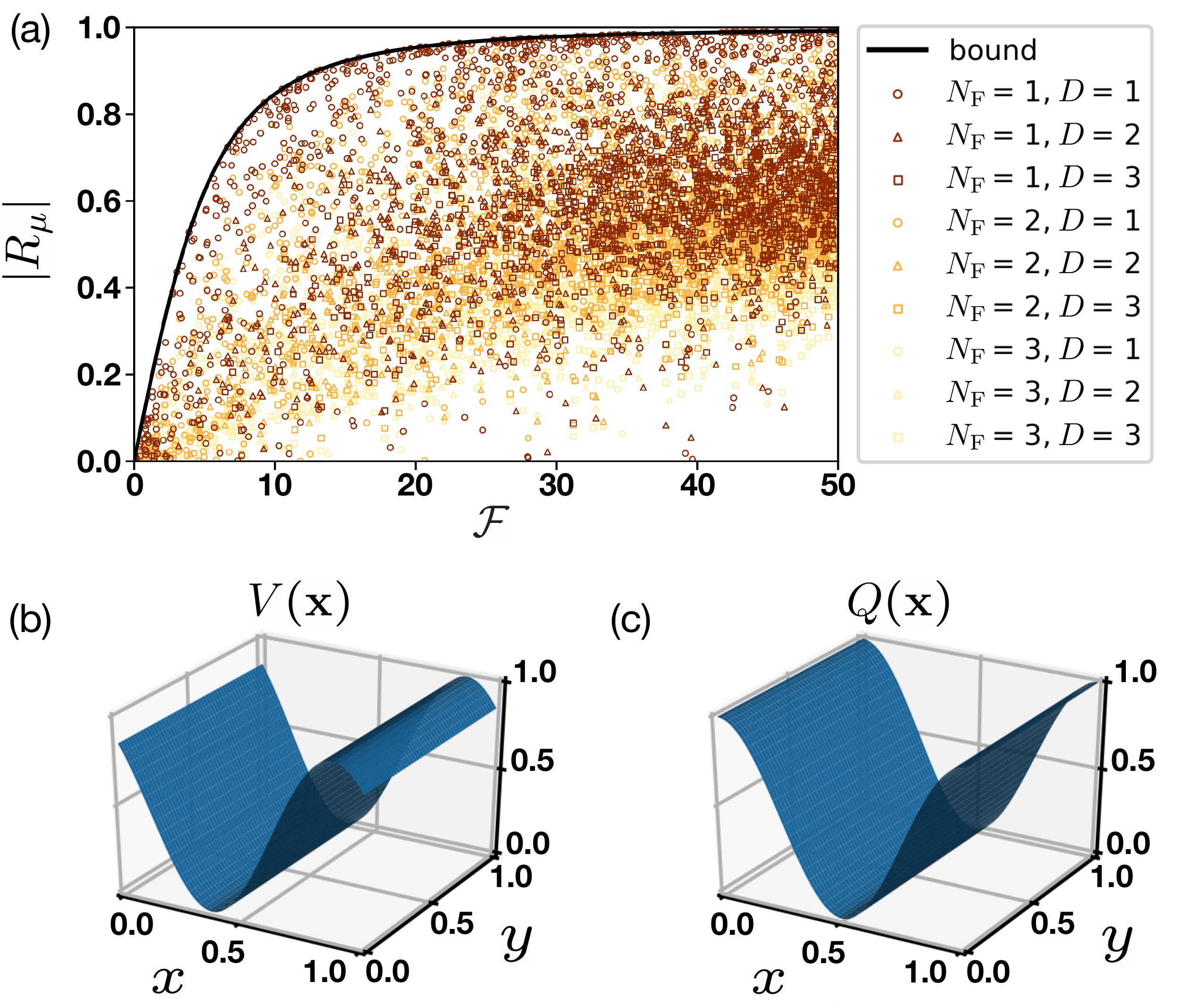}
\caption{Illustration of the general bound \eqref{eq:general_bound}: (a)  Random samples of the response $|R_\mu|$ plotted as a function of the force ${\mathcal F}$.  Samples are generated by choosing $\bf f$ uniformly on $[0,50]^D$, $\hat\mu$ uniformly on $[0,1]^D$, and the Fourier coefficients of $V$ and $Q$ for all allowed modes uniformly on $[0,1]$, while maintaining proper parity so that the corresponding $V$ and $Q$ are real.  These samples are then normalized to keep $\llangle Q^2\rrangle=\llangle V^2\rrangle = 1$.
The colors label the highest Fourier mode $N_{\rm F}$ possible in any direction, while the shapes label the dimension $D$. Each combination of color and shape contains $1000$ data points. (b) \& (c): Example of the optimal $V$ and $Q$ that saturate the bound  \eqref{eq:general_bound} for $D=2$, with ${\bf f}=(f_x,f_y)$, $|f_x|>|f_y|$.}
\label{fig:response_bound_ni}
\end{center}
\end{figure}

Our first main result is the bound
\begin{align}\label{eq:general_bound}
\left|R_\mu \right|\le\sqrt{\frac{\llangle Q^2\rrangle \llangle V^2\rrangle}{1+(2\pi/{\mathcal F})^2}},
\end{align} 
where ${\mathcal F}=\max_j |{f}_j|L$ quantifies how far the system is out of equilibrium. As a sanity check, when ${\mathcal F}=0$, the system is at equilibrium, and the bound is zero, meaning there is no violation of FDT (cf.~\eqref{eq:viol_FDT}). 
We find that equality is reached when $Q$ and $V$ are given by sine waves in one direction (shifted by a phase), and uniform in the orthogonal directions.
One example of such an optimal choice is illustrated in Figs.~\ref{fig:response_bound_ni}(b) and (c) for two dimensions. 
Thus, the system appears most sensitive to slowly varying perturbations that align with the thermodynamic driving.

In our second main result, we specialize to a situation where both observable and conjugate coordinate are the same, $Q=V$.
In this case, we have shown that the response satisfies the tighter inequality
\begin{align}\label{eq:abound}
\left|R_\mu \right|\le\frac{\llangle Q^2\rrangle}{1+(2\pi/{\mathcal F})^2}.
\end{align}
The improvement comes from the denominator being $1+(2\pi/{\mathcal F})^2\ge \sqrt{1+(2\pi/{\mathcal F})^2}$. This is plausible considering that the observable and conjugate coordinate are more constrained.
Despite this improvement, the most sensitive response is again reached for low wave number sine waves.

The third main result is for an even further restricted situation where $Q$ and $V$ are indicator functions. 
It should be stressed that the explicit form, presented below, is based solely on numerical observations for which we have no analytic proof.
To be specific, we require $Q({\bf x})=V({\bf x})=\delta_{\mathcal{S}}({\bf x})$, where $\delta_{\mathcal{S}}(\bf x)$ takes the value one for ${\bf x}\in\mathcal{S}\subseteq\Omega$ and zero otherwise. In this case, we observed that
\begin{align}\label{eq:numerical_bound}
\left|R_\mu \right| &\le \llangle \delta_{\mathcal{S}}^2\rrangle \left[1-\frac{4}{{\mathcal F}}\tanh\left(\frac{{\mathcal F}}{4}\right)\right].
\end{align}
This upper bound corresponds to the response in the case where half of the region $\Omega$ is perturbed (in a way specified below).
That there is a tighter inequality is reasonable, since we are focusing on a more restricted scenario.
Numerical evidence of this improvement will be provided below.

In Fig.\ \ref{fig:response_bound_ni}, we numerically verify inequality \eqref{eq:general_bound} for one, two, and three dimensional driven diffusions in a flat landscape. 
To numerically determine the response, we exploit the translational invariance of the problem to calculate the response of the steady-state distribution in Fourier space, $\delta\pi_{\bf m}$ with ${\bf m}\in{\mathbbm Z}^D$, and then estimate the response from the sum over Fourier coefficients $R_\mu = \sum_{\bf m} \delta \pi_{\bf m}Q_{-{\bf m}}$ (see the discussion leading to \eqref{eq:resFourier} below).
As this is done numerically, the number of Fourier modes that we can keep in this analysis is finite.
We implement this restriction by keeping only modes with ${\bf m}$ below a cut-off $\max_j |m_j| \le N_{\rm F}$, which in this study we take to be one, two or three.
Then, for each combination of $N_{\rm F}$ and $D$, we generate $1000$ random samples by varying the system parameters.  
We observe that in Fig.\ \ref{fig:response_bound_ni} the bound is saturable for every value of driving force.  
Moreover, we observe that potential optima are confined to the lowest space of modes with $N_{\rm F} = 1$, where both observable $Q$ and the conjugate coordinate $V$ are sine waves.
This is why we tend to only observe saturation of the inequality when we restrict our sampling to $N_{\rm F} = 1$, and why our samples fall away from the optimal curve for higher values of $N_{\rm F}$.

\begin{figure}[t]
\begin{center}
\includegraphics[width=.45\textwidth]{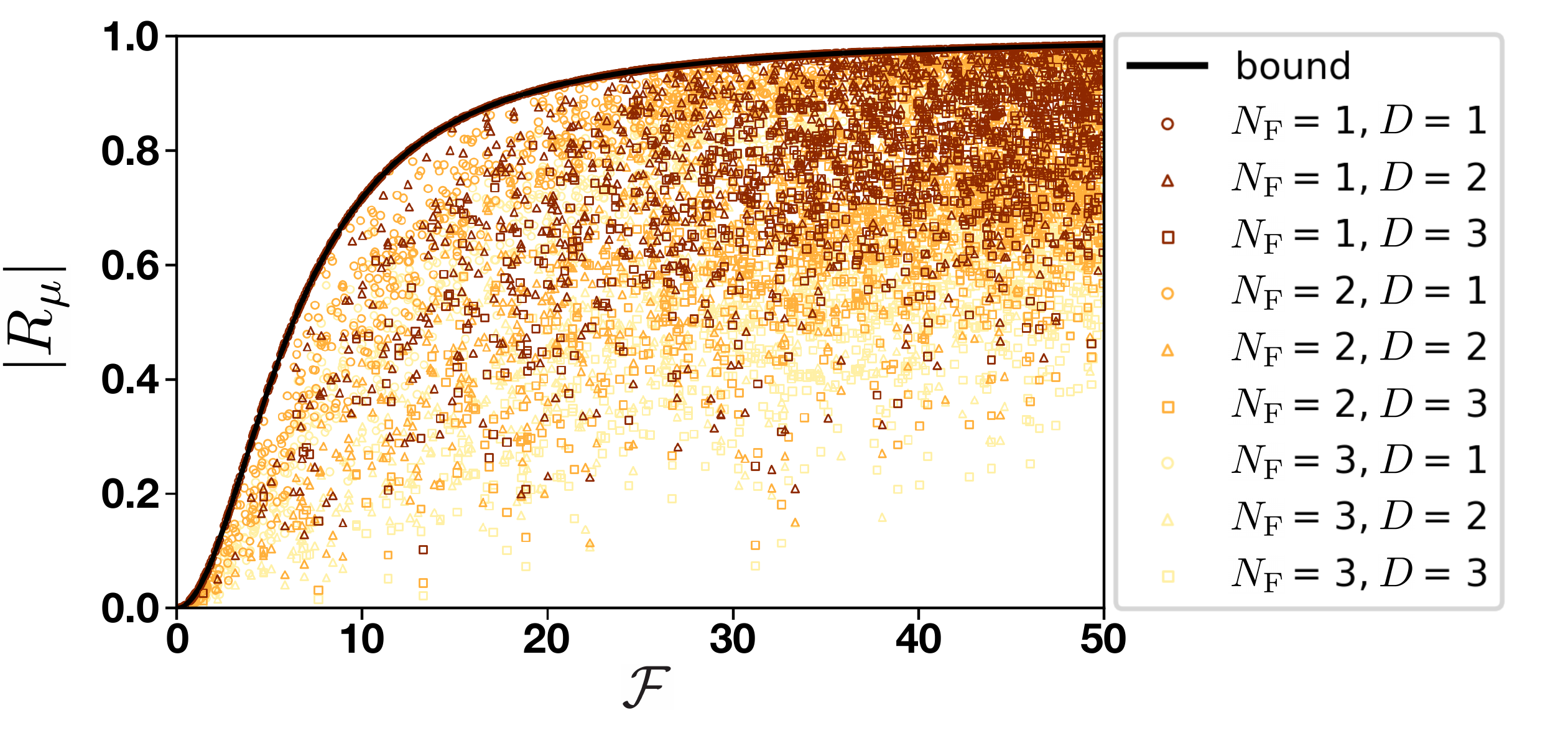}
\caption{Illustration of the bound \eqref{eq:abound}: Random samples of the response $|R_\mu|$ plotted as a function of the force ${\mathcal F}$.  Samples are generated in the same manner as Fig.~\ref{fig:response_bound_ni} except further restricting $V=Q$ and normalizing $\llangle Q^2\rrangle =1$. The colors label the highest Fourier mode $N_{\rm F}$ possible in any direction, while the shapes label the dimension $D$. Each combination of color and shape contains $1000$ data points.
}
\label{fig:response_bound_nc}
\end{center}
\end{figure}
\begin{figure}[t]
\begin{center}
\includegraphics[width=.45\textwidth]{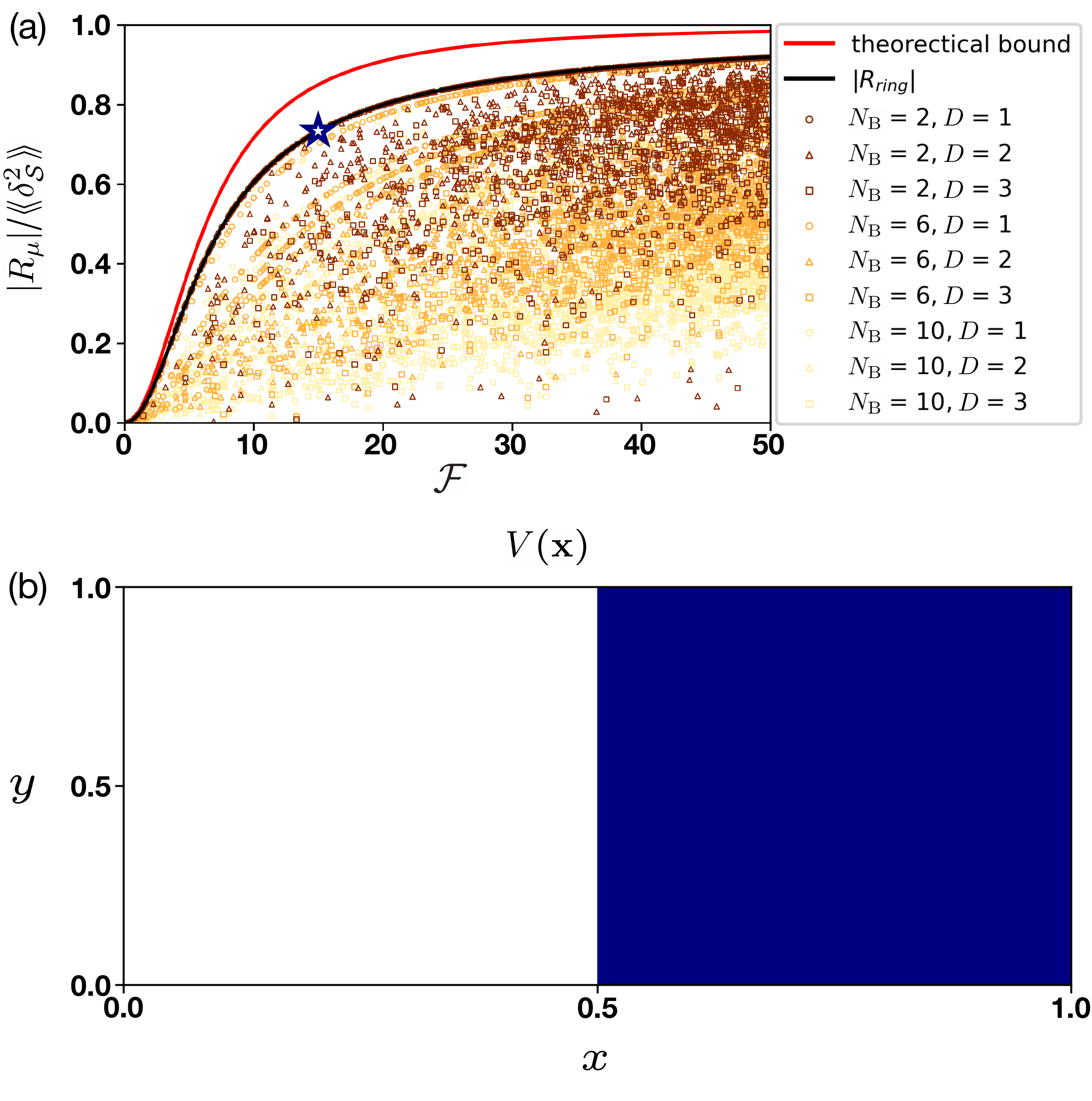}
\caption{Illustration of the bound in \eqref{eq:numerical_bound}: (a) Random sampling of $\hat\mu$, $\bf f$ and perturbation region $\mathcal{S}$. Each combination of color and shape contains $1000$ samples, excluding trivial ones for which $|R_\mu|=\llangle \delta^2_{\mathcal S}\rrangle =0$. We divide $\Omega$ into $N_B^D$ blocks and randomly select each block with probability $0.5$ as the perturbation region. The smaller $N_B^D$, the more probable it is to form an optimal perturbation region to saturate the numerical bound. (b) The perturbation region (blue) that saturates the bound, corresponding to the data point labeled by the star in (a). }
\label{fig:response_bound_uc}
\end{center}
\end{figure}

Our second inequality \eqref{eq:abound} is numerically verified in Fig.~\ref{fig:response_bound_nc}, again in one, two, and three dimensions. Sampling is the same as described for Fig. \ref{fig:response_bound_ni}, except for fixing $V=Q$. All points fall below the predicted limit \eqref{eq:abound} given by the black line. 

Limits to the response when the observable and conjugate coordinate are indicator functions are studied numerically in Fig.~\ref{fig:response_bound_uc}.  
To numerically specify the region ${\mathcal S}$ of the indicator function we divide the entire space $\Omega$ into $N_{\rm B}^D$ equally sized cubic regions we call blocks, with $N_{\rm B}$ the number blocks along each linear dimension.
We can then form a region ${\mathcal S}$ by combining together a collection of these blocks, setting the observable to one on the selected blocks.
For each combination of $D=\{1,2,3\}$ and $N_{\rm B} = \{2,6,10\}$, we generate $1000$ random samples, choosing $\bf f$ and $\hat\mu$ as before.
The perturbation region ${\mathcal S}$ is built up by randomly adding each block to ${\mathcal S}$ with probability $1/2$.
The response-ratio $|R_{\mu}|/\llangle\delta^2_{\mathcal{S}}\rrangle$ is then determined numerically with $N_{\rm F} = 20$ in order to have a Fourier scale finer than the smallest block size.
The data is plotted in Fig.~\ref{fig:response_bound_uc}(a) as a function of force ${\mathcal F}$. 
When $D=1$ and $N_B=6$, there are visible line structures in the plot. This is because there is a small number of different ways of choosing the blocks, which fall on a single line. 
 As $N_B$ increases, these structures remain but are hard to dinstiguish visually.
 For $D\ge 2$, each $\mathcal{F}$ corresponds to infinitely many $\bf f$ so the lines disappear.

The red line is the analytic bound predicted in \eqref{eq:abound}. 
However, it is nowhere saturated.
Instead all sampled responses appear to share a tighter, saturable bound, depicted by the black line.
We visually inspected the samples that appear optimal, like the starred data point in Fig.~\ref{fig:response_bound_uc}(a), and noticed that they consistently all had a perturbation region that occupied half of the full space $\Omega$: for example, the perturbation region for the starred data point is depicted in Fig.\ \ref{fig:response_bound_uc}(b). 
Moreover, we observed that the maxima reached for all the responses is the same for every dimension tested, including 1D.
This suggested to us that we could predict the maximum response possible just from the analytic solution for the 1D ring with perturbation region ${\mathcal S} = [0,1/2]$, 
\begin{align}
|R_\text{ring}|\equiv 1-\frac{4}{{\mathcal F}}\tanh\left(\frac{{\mathcal F}}{4}\right),
\end{align}
which is in fact the black line.

\section{Discussion} We analyzed three different kinetic perturbation schemes of the mobility for homogeneous nonequilibrium diffusions.  For this class of models, we found that the response is not unlimited: it must smoothly approach zero as we near equilibrium and has a maximum arbitrarily far from equilibrium constrained by the size of the fluctuations of the observable and the perturbation's conjugate coordinate. 

Natural extensions would be to have a nonuniform or nondiagonal mobility matrix, or to allow for a anisotropic perturbation. 
Going further and considering arbitrary potentials with a uniform nonconservative force,  $-\nabla U(\bf x)+\bf f$, remain out of reach with the current method, as the Fokker-Planck operator $\mathcal L$ is not diagonal in the Fourier basis. 
Moreover, any bounds for these more general setups would likely take a different form.
Indeed, we have verified that our bounds can be violated in 1D in the presence of a nonuniform potential.
Thus new tools are needed, though the current results suggest progress can be made.
One potential avenue is the recently developed linear-algebraic technique proposed in  \cite{Aslyamov2024}  to tackle single edge perturbations for finite-state Markov chains. 
An extension to diffusion processes would be intriguing, but is far from trivial.
Even when allowing for multi-edge perturbations, taking the diffusive limit presents challenges \cite{PhysRevE.105.L012102}.

\section{Derivation of general variance bound} In this section, we derive the bound on response in terms of variances \eqref{eq:general_bound}.  Without loss of generality, we set the length in each dimension to $L = 1$,to simplify the presentation.

To proceed, we explicitly allow the Fokker-Planck operator to depend on the small external parameter via
\begin{align}
{\mathcal L}_\lambda &= -\nabla\cdot {\hat \mu}(1-\lambda V({\bf x}))\cdot {\bf f}+\nabla\cdot{\hat\mu}(1-\lambda V({\bf x}))\cdot\nabla\\
& \equiv {\mathcal L} +\lambda \delta{\mathcal L}.
\end{align}
with the accompanying steady-state solution $\pi_\lambda$ satisfying ${\mathcal L}_\lambda \pi_\lambda=0$. Since the unperturbed steady-state distribution $\pi$ is uniform, $\pi_\lambda = \pi +\lambda \delta\pi$ is nearly uniform with $\delta\pi = \partial_\lambda \pi_\lambda|_{\lambda=0}$.
With this notation, the response at $\lambda =0$ can be expressed as
\begin{equation}\label{eq:lambdaDerivative}
R_\mu = \int Q({\bf z})\delta  \pi({\bf z})d{\bf z}.
\end{equation}
This expression transfers the problem to determining how the steady-state distribution responds, $\delta \pi$.
A convenient way to obtain this steady-state response is to differentiate the Fokker-Planck equation at $\lambda=0$,
\begin{equation}\label{eq:responseEq}
{\mathcal L} \delta \pi  = -\delta {\mathcal L} \pi =-\nabla V(\bf r)\cdot \hat\mu \cdot\bf f,
\end{equation}
after using $\pi = 1$ is uniform.
This is a partial differential equation for $\delta \pi$~\cite{Martins2023,Khodabandehlou,Aslyamov2024}, which we now proceed to solve.

Because the operator ${\mathcal L}$ is linear in the Fourier basis, a compact solution to \eqref{eq:responseEq} can be found by Fourier transform.
To this end, let us denote the Fourier basis as $e_{\bf m}({\bf r})=e^{i2\pi {\bf m}\cdot{\bf r}}$ for ${\bf m}\in\mathbbm Z^D$.
Because they form an orthonormal basis, $\langle e_{\bf m}, e_{\bf n}\rangle = \int  e_{\bf m}({\bf z}) e^*_{\bf n}({\bf z})d{\bf z}=\delta_{\bf m\bf n}$, any periodic function $G({\bf r})$ can be expanded as
\begin{align}\label{eq:FourierExp}
G(\bf r)&=\sum_{\bf m}G_{\bf m}e_{\bf m}({\bf r}), \quad G_{\bf m}&=\int_\Omega G({\bf r})e^*_{\bf m}({\bf r})d\bf r.
\end{align}
Furthermore, for real $G$, the coefficients satisfy $G^*_{\bf m} = G_{-{\bf m}}$.
This construction is convenient, because once we have determined the Fourier coefficients of the steady-state response $\delta\pi_{\bf m}$, the Fourier expansions in \eqref{eq:FourierExp} can be substituted into \eqref{eq:lambdaDerivative} to obtain the response formula
\begin{equation}\label{eq:resFourier}
R_\mu = \sum_{\bf m}  \delta\pi_{\bf m} Q_{-\bf m}.
\end{equation}

Now to solve \eqref{eq:responseEq}, we expand both sides in the Fourier basis
\begin{equation}\label{eq:FPfourier}
\sum_{\bf m}\delta\pi_{\bf m}{\mathcal L} e_{\bf m}({\bf r})=-\sum_{\bf m}(\delta{\mathcal L}\pi)_{\bf m}e_{\bf m}({\bf r}),
\end{equation}
We first note that in this basis the Fokker-Planck operator is diagonal,
\begin{align}
\mathcal L e_{\bf m}({\bf r})=l_{\bf m}e_{\bf m}(\bf r),
\end{align} 
with eigenvalues $l_{\bf m}=-4\pi^2({\bf m}\cdot\hat\mu\cdot{\bf m})-2\pi i{\bf m}\cdot\hat\mu\cdot \bf f$.
Next, we evaluate the right hand side using \eqref{eq:responseEq}, which reads
\begin{align}
(\delta {\mathcal L}\pi)_{\bf m}&=-\int_\Omega[\nabla V({\bf r})\cdot \hat\mu \cdot{\bf f}] e^*_{\bf m}({\bf r})d{\bf r}, 
\\
&=-2\pi i V_{\bf m}({\bf m}\cdot\hat\mu\cdot \bf f),  
\end{align}

Combining these observations with the orthogonality of the Fourier basis leads to a series of uncoupled linear equations for $\delta\pi_{\bf m}$, which can be solved
\begin{align}\label{eq:response_m} 
\delta\pi_{\bf m}=-\frac{1}{l_{\bf m}} (\delta {\mathcal L}\pi)_{\bf m}= \frac{2\pi i V_{\bf m}({\bf m}\cdot\hat\mu\cdot {\bf f})}{-4\pi^2({\bf m}\cdot\hat\mu\cdot{\bf m})-2\pi i{\bf m}\cdot\hat\mu\cdot {\bf f}}
\end{align}
for $ {\bf m}\neq\bf 0$ with $\delta\pi_{\bf 0}=0$ due to probability conservation.
Now substituting this solution into \eqref{eq:resFourier} leads to our desired starting point for deriving bounds, 
\begin{align}\label{eq:R_SQ_series2}
R_\mu&=-\sum_{\bf m\neq\bf 0}\lambda_{\bf m}V_{\bf m}Q_{-\bf m},
\end{align}
with
\begin{align}
\lambda_{\bf m}=\frac{\bf m\cdot\hat\mu\cdot{\bf f}}{{\bf m}\cdot\hat\mu\cdot{\bf f}-2\pi i{\bf m}\cdot\hat\mu\cdot{\bf m}}.
\end{align}

We now bound the maximum of \eqref{eq:R_SQ_series2} over all $V_{\bf m}$ and $Q_{\bf m}$, with their variances constrained,
\begin{align}
\llangle V^2\rrangle = \sum_{\bf m\neq 0}|V_{\bf m}|^2, \quad
\llangle Q^2\rrangle = \sum_{{\bf m}\neq 0}|Q_{{\bf m}}|^2.
\end{align}
To proceed, we introduce re-weighted Fourier coefficients ${\tilde V}_{\bf m} = \sqrt{\lambda_{\bf m}}V_{\bf m}$ and ${\tilde Q}_{\bf m} = \sqrt{\lambda_{-{\bf m}}}Q_{\bf m}$, allowing us to apply the Cauchy-Scwharz inequality to \eqref{eq:R_SQ_series2}:
\begin{align}
\left|R_\mu\right|&=\left|\sum_{\bf m\neq\bf 0}{\tilde V}_{\bf m}{\tilde Q}^*_{\bf m}\right| \\
&\le \sqrt{\left(\sum_{\bf m\neq\bf 0}|\lambda_{\bf m}||V_{\bf m}|^2\right)\left(\sum_{\bf m\neq\bf 0}|\lambda_{\bf m}||Q_{\bf m}|^2\right)}\\
&\le \max_{\bf m\neq 0}|\lambda_{\bf m}|\sqrt{\llangle V^2\rrangle \llangle Q^2\rrangle}
\label{eq:CS_response}
\end{align}
 All that is left is to bound the magnitude of $\lambda_{\bf m}$.
 Let the real and imaginary parts of $\lambda_{\bf m}$ be $\rho_{\bf m}$ and $\sigma_{\bf m}$: $\lambda_{\bf m}=\rho_{\bf m} +i\sigma_{\bf m}$. Then the squared magnitude of $\lambda_{\bf m}$, which here equals its real part $\rho_{\rm m}$, can be written as
\begin{align}
|\lambda_{\bf m}|^2=\rho^{2}_{\bf m}+\sigma^{2}_{\bf m}=\rho_{\bf m}=\frac{1}{1+4\pi^2\left(\frac{{\bf m}\cdot\hat\mu\cdot{\bf m}}{\bf m\cdot\hat\mu\cdot{\bf f}}\right)^2}.
\end{align}
To bound $|\lambda_{\bf m}|^2$, we study the quantity in the denominator
\begin{align}
Y_{\bf m}&=\left|\frac{{\bf m}\cdot\hat\mu\cdot{\bf f}}{{\bf m}\cdot\hat\mu\cdot{\bf m}}\right|=\left|\sum_{i|m_i\neq 0} w_i \frac{f_i}{m_i}\right|,
\end{align}
where $w_i$ are non-negative weights 
\begin{align}
w_i=\frac{\mu_i m_i^2}{\sum_{i|m_i\neq 0} \mu_i m^2_i}
\end{align}
satisfying $\sum_{i|m_i\neq 0} w_i=1$, with the sum extending over all $i$ such that $m_i\neq 0$. 
 By introducing the weights $w_i$, we can treat $Y_{\bf m}$ as the absolute value of the average of $f_i/m_i$.
Now, by varying ${\bf m}$, holding $\hat\mu$ and ${\mathcal F}$ fixed, we can vary the  $w_i$ in order to find the largest attainable value of $Y_{\bf m}$.
But as an average, that value can be no larger than the 
 the maximal $f_i/m_i$, which is $\mathcal{F}$. 
With $Y_{\bf m}$ bounded by $\mathcal{F}$, we deduce 
\begin{align}
|\lambda_{\bf m}|^2=\rho_{\bf m}&\le\frac{1}{1+(2\pi/{\mathcal F})^2}.\label{eq:rho_bound}
\end{align}
Substitution of this bound into \eqref{eq:CS_response}, we arrive at the desired result \eqref{eq:general_bound}.

To reach our bound we have two inequalities to saturate.
First, the maximum in \eqref{eq:rho_bound} is reached when ${\bf m}\neq{\bf 0}$ satisfies the following: for the $i\in\text{argmax}_{j}|f_j|$, $m_i\in\{1,0,-1\}$, and for the $i\notin\text{argmax}_{j}|f_j|$, $m_i=0$; besides for any $i\neq j$, $f_i f_j m_i m_j\geq 0$. Here we have considered the case where $|f_i|$ could be the same as $|f_j|$.
Next, the condition for equality in the Cauchy-Schwarz inequality is that there is an ${\bf m}$-independent constant $c$ such that $\sqrt{\lambda_{\bf m}}V_{\bf m}= c \sqrt{\lambda_{-{\bf m}}}Q_{\bf m}$.
Substituting in the polar form of $\lambda_{\bf m} = r_{\bf m}e^{i \phi_{\bf m}}$, we see that equality in our bound requires the observable and conjugate coordinate to be equal up to a phase-shift and rescaling 
\begin{equation}
V_{\bf m} = c e^{-i\phi_{\bf m}} Q_{\bf m}.
\end{equation}

\section{Derivation $Q=V$ bound} Since $Q=V$, the response \eqref{eq:R_SQ_series2} takes the form
\begin{align}\label{eq:u_c_response}
R_{\mu}
&=-\sum_{{\bf m}\neq{\bf 0}}\rho_{\bf m}|Q_{\bf m}|^2,
\end{align}
where we have taken only the real part of \eqref{eq:R_SQ_series2} as $R_\mu$ must be real.
As each term is positive, the absolute value of this sum can be bounded as 
\begin{align}\label{eq:u_c_response}
|R_{\mu}|
&\le\max_{{\bf m}\neq{\bf 0}}\rho_{\bf m}\sum_{{\bf m}\neq{\bf 0}}|Q_{\bf m}|^2= \max_{{\bf m}\neq{\bf 0}}\rho_{\bf m}\llangle Q^2\rrangle,
\end{align}
which using \eqref{eq:rho_bound} immediately leads to the desired bound.
The saturation condition immediately follows: $Q=V$ only have nonzero Fourier components in $\text{argmax}_{\bf m}\rho_{\bf m}$. 

\section{Maximum response for a flat diffusion in 1D} 
In this section, we calculate the conjectured limit in \eqref{eq:numerical_bound} depicted as the black line in Fig.~\ref{fig:response_bound_uc}.
This bound is given by the response of a diffusion process on a unit-length ring with a constant mobility $\mu(z)=\mu$,  constant driving force $f>0$, due to a perturbation of the mobility in region $[0,1/2]$:
\begin{align}
R_\mu = \int_0^{\frac{1}{2}}\int_0^{\frac{1}{2}}\frac{\delta\pi_\text{ring}(x)}{\delta\ln\mu(z)}dxdz
\end{align}
The closed-form expression for the integrand is given in \cite{PhysRevE.105.L012102}, which we reproduce here. Defining the functions
\begin{align}\label{eq:S}
 &S(x',x)=e^{-f(x'-x)}\left[e^{-f}\Theta(x-x')+\Theta(x'-x)\right],\\
 \label{eq:N}
 &\mathcal N=\int_0^1\int_0^1 S(x',x)\ dx' dx = (1-e^{-f})/f,
\end{align}
where $\Theta(z)$ is the Heaviside step function that is one for $z>0$ and zero otherwise, we can write the steady-state response as
\begin{equation}
\frac{\delta \pi_{\rm{ring}}(x)}{\delta\ln\mu(z)} = -\frac{1}{{\mathcal N}}{S}(z,x)+\frac{1}{\mathcal N}\int_0^1 S(z,y) dy.
\end{equation}
This expression can be evaluated analytically, which gives the bound on the response
\begin{align}
|R_\mu|\le\frac{1}{4}-\frac{1}{f}\tanh\left(\frac{f}{4}\right).
\end{align}
In this case $\llangle \delta_{\mathcal{S}}^2\rrangle=1/4$ also reaches its maximum.
Thus, we find the analytical expression of the saturable numerical bound presented in \eqref{eq:numerical_bound}.

\acknowledgments
This material is based upon work supported in part by the National Science Foundation under Grant No. 2142466 and by the Alfred P. Sloan Foundation under grant G-2022-19440. H.-M.C. was supported by a KIAS Individual Grant (PG089401) at Korea Institute for Advanced Study.
\bibliographystyle{eplbib}
\bibliography{Bib.bib}

\end{document}